\documentclass[conference]{IEEEtran}
\IEEEoverridecommandlockouts
\usepackage{cite}
\usepackage{amsmath,amssymb,amsfonts}
\usepackage{algorithm}
\usepackage{algpseudocode}
\usepackage{comment}
\usepackage[pdftex,colorlinks=true,urlcolor=blue,citecolor=black,anchorcolor=black,linkcolor=black]{hyperref}
\usepackage[capitalise]{cleveref}
\usepackage{graphicx}
\usepackage{textcomp}
\usepackage{xcolor}
\def\BibTeX{{\rm B\kern-.05em{\sc i\kern-.025em b}\kern-.08em
    T\kern-.1667em\lower.7ex\hbox{E}\kern-.125emX}}
\begin{document}

\title{Estimating Treatment Effects Using Costly Simulation Samples from a Population-Scale Model of Opioid Use Disorder\\
}

\author{\IEEEauthorblockN{Abdulrahman A. Ahmed}
\IEEEauthorblockA{\textit{Dept. of Industrial Engineering} \\
\textit{University of Pittsburgh}\\
Pittsburgh, PA, USA \\
aba173@pitt.edu}
\and
\IEEEauthorblockN{M. Amin Rahimian}
\IEEEauthorblockA{\textit{Dept. of Industrial Engineering} \\
\textit{University of Pittsburgh}\\
Pittsburgh, PA, USA \\
rahimian@pitt.edu}
\and
\IEEEauthorblockN{Mark S. Roberts}
\IEEEauthorblockA{\textit{Dept. of Health Policy and Management} \\
\textit{University of Pittsburgh}\\
Pittsburgh, PA, USA \\
mroberts@pitt.edu}
}

\maketitle

\begin{abstract}
Large-scale models require substantial computational resources for analysis and studying treatment conditions. Specifically, estimating treatment effects using simulations may require a lot of infeasible resources to allocate at every treatment condition. Therefore, it is essential to develop efficient methods to allocate computational resources for estimating treatment effects. Agent-based simulation allows us to generate highly realistic simulation samples. FRED (A Framework for Reconstructing Epidemiological Dynamics) is an agent-based modeling system with a geospatial perspective using a synthetic population constructed based on the U.S. census data. Given its synthetic population, FRED simulations present a baseline for comparable results from different treatment conditions and treatment conditions. In this paper, we show three other methods for estimating treatment effects. In the first method, we resort to brute-force allocation, where all treatment conditions have an equal number of samples with a relatively large number of simulation runs. In the second method, we try to reduce the number of simulation runs by customizing individual samples required for each treatment effect based on the width of confidence intervals around the mean estimates. In the third method, we use a regression model, which allows us to learn across the treatment conditions such that simulation samples allocated for a treatment condition will help better estimate treatment effects in other (especially nearby) conditions. We show that the regression-based methods result in a comparable estimate of treatment effects with less computational resources. The reduced variability and faster convergence of model-based estimates come at the cost of increased biased, and the bias-variance trade-off can be controlled by adjusting the number of model parameters (e.g., including higher-order interaction terms in the regression model). 
\end{abstract}


\section{Introduction}

Estimating treatment effects for large-scale models is hard. In reality, this may be an expensive and time-consuming task. Cranmer et al. \cite{cranmer2020frontier} discuss possible machine learning techniques for inference when (simulation) models become more complex. Agent-based simulation appears as a solution when conducting experiments is infeasible and allows us to utilize computational power to circumvent these obstacles. Shea et al. \cite{shea2023covid} use agent-based simulation to evaluate treatment effects for epidemic outbreaks (e.g., COVID-19). 
Running agent-based simulation over large populations requires a lot of computational resources, and it becomes more challenging when there are multiple treatment conditions to evaluate and optimize. Hence different techniques are proposed to tackle the costly computations of population-scale simulation models.

Moreover, this problem is similar to other problems like  Bayesian Optimization (BO), where the objective function evaluation is costly, and we have few chances to get the extreme value. Frean and Boyle \cite{frean2008nn} use BO to learn the weights of 
 a neural network controller to balance two vertical poles simultaneously. Another area related to this problem is active learning, where the machine learns by as few labeled data as possible with little assistance needed to continue the task (i.e., no added labeling by a human). This circumvents the cost of labeling large amounts of data \cite{settles2009active}. The problem also incorporates the concept of exploitation vs. exploration, where we can run a simulator by changing the parameter $\theta$ and exploiting the information that we get from simulated samples to lead us to where to explore next. This problem is also related to Bayesian experimental design, where a utility function is updated iteratively to improve information from outcomes \cite{chaloner1995bayesian}. 
Multi-armed bandit (MAB) is another related problem area, where the goal is to maximize the gain/reward by choosing limited options out of a set of alternatives. MAB also exhibits the exploration-exploitation trade-off, i.e., whether to keep selecting the same arms or to explore potential gains in other arms \cite{sutton2018reinforcement}. Lastly, our methods touch on the classical problem of bias-variance trade-off in model selection, where the goal is to strike a desirable balance between the two often opposing sources of error.

This paper is structured into five sections. In \cref{sec:concept}, we give a brief introduction to the FRED simulation software and details about how FRED works, while in its second part, we discuss the OUD model that we will use to apply treatment conditions. In \cref{sec:method}, we will demonstrate different methods we used to estimate treatment effects. In \cref{sec:result}, we will discuss the results of our study and its public health implications. Finally, in \cref{sec:conc}, we provide concluding remarks and give future work directions.

\section{Preliminary concepts} \label{sec:concept}
\subsection{FRED simulation framework}
FRED (Framework for Reconstructing Epidemiological Dynamics) is an agent-based, open-source simulation software that is developed to simulate the temporal and spatial behaviors of epidemics. Public Health Dynamics Laboratory (PHDL) in the University of Pittsburgh School of Public Health is behind the development of the FRED software. Originally, FRED was designed to study the dynamics of an epidemic. However, FRED has shown broader potential for large-scale population studies that could help in providing a better understanding of public health treatment conditions and policies. One of the strong points of FRED is that it has a synthetic population that is accurately based on the US Census Bureau's public use microdata files and Census aggregated data \cite{guclu2016agent}.

\subsubsection{Synthetic Population}
Every individual in FRED is represented explicitly in a designated geographic area. FRED utilizes the US synthetic population database from RTI International \cite{rti}, where the synthetic population contains detailed geographically allocated categories. In the synthetic population in FRED, each agent has an assigned household. Also, there are assigned agents for each facility. Each household, school, and workplace is set to a specific region \cite{grefenstette2013fred}. 

\subsubsection{Discrete-time simulation}
At every simulation stage, the agent communicates with other agents who are likely to share the same daily occupation. For example, agents in the same school interact with the same colleagues on a daily basis. Moreover, suppose the agent is infected and interacts with a susceptible agent. In that case, there is a chance for disease transmission from the infected agent to the susceptible one.

\subsection{OUD model}
The Opioid Use Disorder (OUD) model is developed to understand the OUD epidemic in the U.S., where opioids are the leading cause of drug overdose deaths (that includes prescription opioids, heroin, and synthetic opioids \cite{CDCOpioids}). Jalal et al. \cite{jalal2018opioiddynamics} studied the epidemic dynamics over the last 40 years and reached the conclusion that the present opioid overdose deaths wave is a part of a long trend that is undergoing over several decades, hence, stressing the importance of studying the epidemic dynamics. 
The OUD model that we use in this paper was developed by the Public Health Dynamics Laboratory at the University of Pittsburgh, based on data provided by the Centers for Disease Control and Prevention (CDC) as a part of their funded research. OUD model has explicitly defined transition probabilities between different states and dwell times for each agent at different states. The OUD model was simulated over the synthetic population of Allegheny County, PA. 

\noindent The model was simulated for the period between Jan 1st, 2016, to Dec 31st, 2017. The simulation was conducted over this specific time frame because state transitions of the OUD model were calibrated with the real data for that time frame.

\begin{figure}[h]
    \centering
    \includegraphics[scale=0.3]{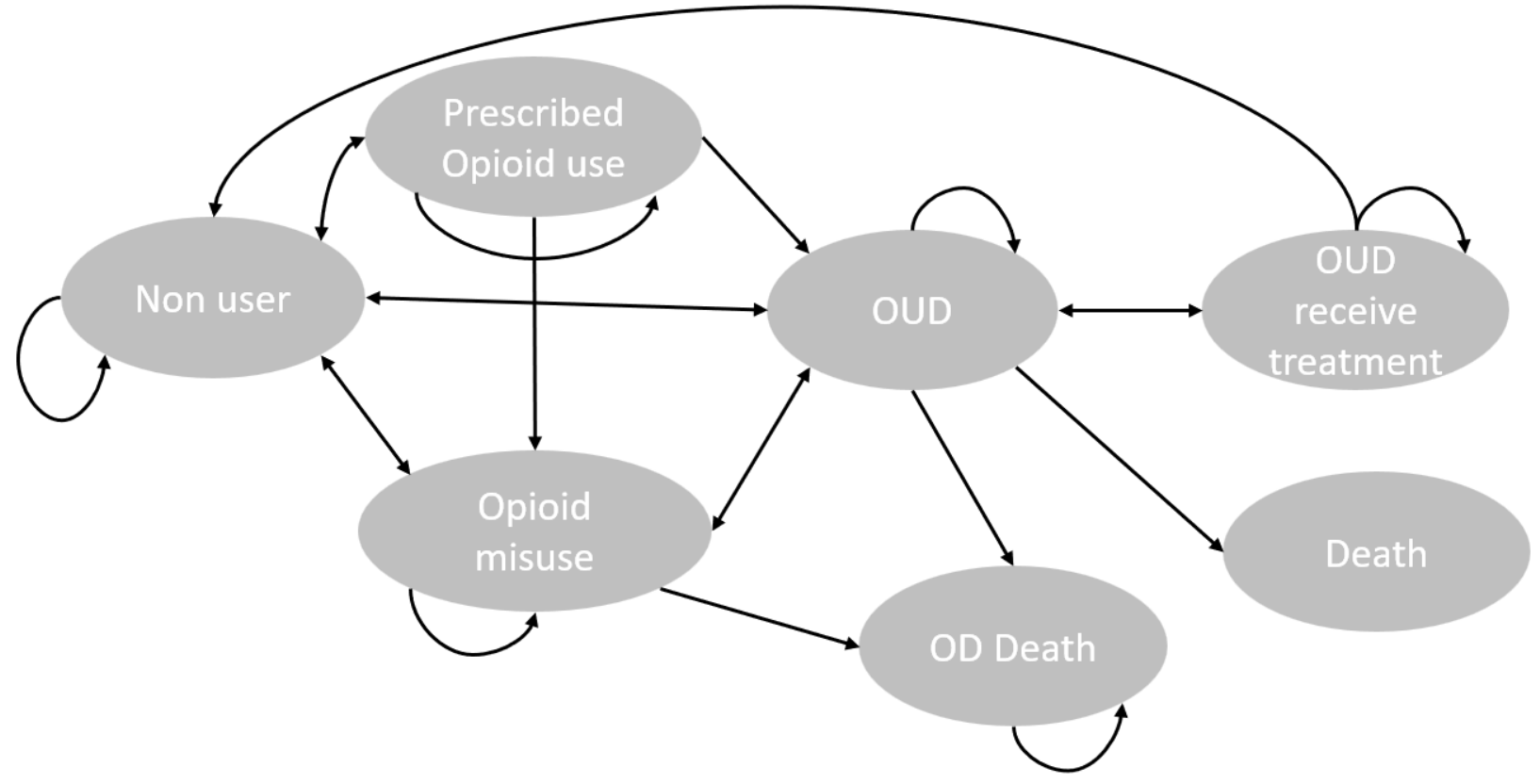}
    \caption{State transition diagram for the OUD model.}
    \label{fig:oudFlow}
\end{figure}

\subsection{Bayesian Optimization}
Although in this study we will not use Bayesian Optimization (BO) directly, it is strongly related to the discussed methods and worth pointing out briefly. BO is a powerful tool for finding the extreme point of a function that is expensive to evaluate. Its best use case is where one cannot obtain a closed-form expression for an objective function but only can obtain observations of this objective function \cite{brochu2010botutorial}. BO contains the concept of prior belief about the function (hence the Bayesian part) and the trade-off between exploitation and exploration of the search space. Similarly, our problem is to maximize the information we get over treatment condition space while using as few resources as possible due to the computational cost of evaluating treatment effects in population-scale agent-based models. 

\section{proposed methods} \label{sec:method}
In this paper, we will use the OUD model to conduct our experiment to estimate the effect of different treatments. The two factors studied in this case are Buprenorphine and Naloxone. Buprenorphine is a medication used as a treatment for OUD, and Naloxone is a medication used as an opioid overdose antidote, i.e., it can reverse the effect of an opioid overdose. An increase in the availability of Buprenorphine will increase the probability of agents moving from OUD to treatment. In contrast, an increase in the amount of Naloxone will decrease the number of overdose deaths or OD Deaths (recall figure \ref{fig:oudFlow}). We selected these two factors as they are highly effective in increasing individuals in treatment from OUD and decreasing the number of OD deaths compared to other possible factors. Each factor has five levels which we will call (A, B, C, D, E, and F) for Naloxone levels and (a, b, c, d, e, and f) for Buprenorphine levels, which constitute 25 treatment conditions (combinations of two factors in five levels each). The 25 conditions can be grouped into five sets by fixing the Naloxone level. In this study, we will report the results of the first two sets (i.e., the first ten treatment conditions).

\subsection{Brute-force method}
The brute-force method allocates an equal number of samples to each treatment condition. Although this method is definite in providing a solid estimate for each treatment condition, it requires a lot of computational resources to reach that result. Moreover, it has an embedded assumption that all treatment conditions have the same uncertainty, which is not valid, as some treatment conditions may require more samples to reach the same confidence interval (CI) width as other treatment conditions. 

\subsection{Greedy method}
Estimates of treatment effects are not created equal. The Greedy method is built on the assumption that some treatment conditions may require more samples than other treatment conditions to reach the same CI width. At first, the method will do an initial equal sample sweep and by conducting a fixed number of simulation runs. Afterwards, the allocations depend on the width of the CIs, and the treatment condition with the widest CI will receive the next batch of samples. Algorithm \ref{alg:cap2} shows the procedure we used to implement the greedy method. Table \ref{tab:tab2} shows the mean and CI width for each treatment condition. 
\begin{algorithm}
\caption{Greedy method}\label{alg:cap2}
\begin{algorithmic}
\State Do initial $n$ simulation runs for each treatment condition
\State initialize flag=0
\While{flag $\neq$ 1}
    \For{j:=1}{ 10}
    \State Get the largest CI between treatment condition estimation
    \State Assign largest CI to variable $max$
    \EndFor
    
    \If{$max<$ 5}
        \State flag=1
    \Else
        \State get $n$ simulation runs for treatment condition with $max$
    \EndIf
\EndWhile
\end{algorithmic}
\end{algorithm}

\begin{table}[h]
    \centering
    \begin{tabular}{|c|c|c|c|}
    \hline
        TC & Mean & CI width & number of runs  \\ \hline
        Aa & 2390.35 & 3.94 & 2450 \\ \hline
        Ab & 2375.09 & 3.93 & 2450 \\ \hline
        Ac & 2361.39 & 3.97 & 2300 \\ \hline
        Ad & 2346.5 & 3.94 & 2350 \\ \hline
        Ae & 2331.83 & 3.93 & 2200 \\ \hline
        Ba & 2383.39 & 3.95 & 2350 \\ \hline
        Bb & 2368.21 & 3.96 & 2300 \\ \hline
        Bc & 2351.93 & 3.95 & 2300 \\ \hline
        Bd & 2339.26 & 3.96 & 2300  \\ \hline
        Be & 2323.23 & 3.97 & 2350 \\
        \hline
    \end{tabular}
    \vspace{0.25cm}
    \caption{Greedy results for different treatment conditions given that method stop as the maximum width $<$4, where the total number of samples is 23350. \vspace{-10pt}}
    \label{tab:tab2}
\end{table}

\subsection{Model-based greedy method}
What if we can use samples simulated for a specific treatment condition to learn about other neighboring treatment conditions? In this section, we use the greedy method to allocate simulation samples to estimate the  parameters of a linear regression model based on all of the simulation samples conducted in all ten treatment conditions. We will refer to this method as ``model-based greedy''. For example, getting samples for treatment condition five will not only tell us about CI over treatment condition five but also provide information about treatment condition four and treatment condition six. Equation \eqref{eq:regressionWITHinteraction} shows the regression for the treatment effect $y$ given the level of Buprenorphine $x_1$ and level of Naloxone $x_2$:
\begin{equation}
    y=\beta_0+\beta_1 x_1+\beta_2 x_2+ \beta_3 x_1 x_2
    \label{eq:regressionWITHinteraction}
\end{equation}
Algorithm \ref{alg:cap3} shows our implementation of the model-based greedy approach. Practically the method was evaluated using the CI around each treatment condition where each time a new batch is added to the selected treatment condition until all treatment conditions' CIs are below a predefined threshold. It is worth noting that our assumption of a straightforward model (i.e., the linear regression model) may not be the best fit for the problem but helps us estimate the treatment effects in a situation where computational resources are costly.

\begin{algorithm}
\caption{Model-based greedy method}\label{alg:cap3}
\begin{algorithmic}
\State Do initial $n$ simulation runs for each treatment condition
\State Get initial values for regression model parameters
\State Define $threshold$ as the threshold for acceptable error
\State Initialize flag=0
\While{flag $\neq$ 1}
    \State Optimize regression model parameters
    \State Do $n$ simulation runs for each treatment condition
    \State Calculate $e$ as the error between regression model parameters and samples
    \If{$e< threshold$}
        \State flag=1
    \EndIf
\EndWhile
\end{algorithmic}
\end{algorithm}

\begin{table}[h]
\begin{tabular}{|l|lll|lll|}
\hline
             & \multicolumn{3}{c|}{Model-based greedy}                                                  & \multicolumn{3}{c|}{Model-based without interaction}                                                \\ \hline
TC & \multicolumn{1}{l|}{Mean}    & \multicolumn{1}{l|}{CI width} & \# runs & \multicolumn{1}{l|}{Mean}    & \multicolumn{1}{l|}{CI width} & \# runs \\ \hline
Aa            & \multicolumn{1}{l|}{2390.12} & \multicolumn{1}{l|}{3.96}     & 1200       & \multicolumn{1}{l|}{2388.44} & \multicolumn{1}{l|}{4.5}      & 650        \\ \hline
Ab            & \multicolumn{1}{l|}{2375.02} & \multicolumn{1}{l|}{2.82}     & 900        & \multicolumn{1}{l|}{2374.1}  & \multicolumn{1}{l|}{3.6}      & 500        \\ \hline
Ac            & \multicolumn{1}{l|}{2359.9}  & \multicolumn{1}{l|}{2.4}      & 900        & \multicolumn{1}{l|}{2359.52} & \multicolumn{1}{l|}{3.2}      & 500        \\ \hline
Ad            & \multicolumn{1}{l|}{2344.82} & \multicolumn{1}{l|}{3.01}     & 900        & \multicolumn{1}{l|}{2344.95} & \multicolumn{1}{l|}{3.4}      & 500        \\ \hline
Ae            & \multicolumn{1}{l|}{2329.72} & \multicolumn{1}{l|}{4.22}     & 1550       & \multicolumn{1}{l|}{2330.37} & \multicolumn{1}{l|}{4.1}      & 500        \\ \hline
Ba            & \multicolumn{1}{l|}{2384.63} & \multicolumn{1}{l|}{2.58}     & 900        & \multicolumn{1}{l|}{2383.25} & \multicolumn{1}{l|}{3.3}      & 500        \\ \hline
Bb            & \multicolumn{1}{l|}{2369.89} & \multicolumn{1}{l|}{1.83}     & 900        & \multicolumn{1}{l|}{2368.68} & \multicolumn{1}{l|}{2.4}      & 500        \\ \hline
Bc            & \multicolumn{1}{l|}{2354.78} & \multicolumn{1}{l|}{1.53}     & 900        & \multicolumn{1}{l|}{2354.1}  & \multicolumn{1}{l|}{2.1}      & 500        \\ \hline
Bd            & \multicolumn{1}{l|}{2339.67} & \multicolumn{1}{l|}{1.9}      & 900        & \multicolumn{1}{l|}{2339.53} & \multicolumn{1}{l|}{2.7}      & 1000
\\ \hline
Be            & \multicolumn{1}{l|}{2323.71} & \multicolumn{1}{l|}{2.3}      & 900        & \multicolumn{1}{l|}{2324.95} & \multicolumn{1}{l|}{2.7}      & 500    
\\ \hline
             & \multicolumn{1}{l|}{}        & \multicolumn{1}{l|}{Total=}   & 9950       & \multicolumn{1}{l|}{}        & \multicolumn{1}{l|}{Total=}   & 5650       \\ \hline
\end{tabular}
\vspace{3pt}
 \caption{Model-based greedy and Model-based greedy without interaction mean results for the first ten treatment conditions effects and corresponding CI width where the third column shows the number of runs required to reach that estimate. \vspace{-10pt}}
    \label{tab:tab3}
\end{table}

\subsection{Model-based greedy method without interaction}
Recall the bias-variance dilemma of model selection where we can trade less variability for more bias by using simpler models that are easier to estimate (more precise but less accurate). To demonstrate this concept, we remove the interaction term from \eqref{eq:regressionWITHinteraction} and use the same model-based greedy method with a simpler model to see how this could affect the estimates for the treatment effects and their sample size requirements. Equation \eqref{eq:regressionWITHOUTinteraction} shows the regression equation for treatment condition $y$ given the level of Buprenorphine $x_1$ and level of Naloxone $x_2$, without the interaction term.
\begin{equation}
    y=\beta_0+\beta_1 x_1+\beta_2 x_2
    \label{eq:regressionWITHOUTinteraction}
\end{equation}
We use the same algorithm as before for model-based greedy, except for the simplification in the regression model.


\section{Results and Discussion} \label{sec:result}
To compare the effect of each treatment, we selected OD Deaths as the target for our experiment. The treatment condition that gets the lower OD Deaths is better. Although the first method used relatively more extensive simulation runs, its estimates are similar to the other models, showing that the latter models could save computational resources critical for larger simulations (e.g., entire PA state or nationwide) with more factors and levels to intervene at (i.e., exponentially more treatment conditions). Specifically, model-based greedy showed that it got most of the estimates right (with narrower CI) with almost half of the simulation runs compared to the greedy method. This also implies that the saving in the number of simulation runs will increase as the number of treatment conditions increases.
Moreover, using the model-based greedy with a simpler regression model (with no interaction terms) showed the potential to improve the bias-variance trade-off of model selection.  The estimation results with model-based greedy without interaction in table \ref{tab:tab3} showed that we can achieve performance on par with the more complex model with a small bias in estimating the mean of treatment effects and almost half the sample size.


\section{Conclusion} \label{sec:conc}
Estimating treatment effects in large-scale models is a complicated and computationally exhaustive problem. In this paper, we showed that by using simple techniques, we can save on computational resources by estimating the same quantities with fewer simulation runs. We demonstrated three methods for this: 1) the brute-force method, allocating simulation runs equally across treatment conditions, 2) the greedy method that improves on brute-force by allocating to the less precise conditions first, and 3) model-based greedy, which attempts to reduce sample size requirements by assuming a regression model for the effect size across treatment conditions. Finally, we demonstrated that even a simple model-based greedy method without interaction terms can achieve comparable performance with even fewer samples while sacrificing some accuracy (i.e., bias-variance trade-off).
This work can be extended by: 1) devising better allocation strategies that improve on greedy by considering the effect of allocations on the model estimates across all conditions (e.g., using Bayesian optimization), and 2) improving the bias-variance trade-off of model selection using more expressive model classes to better approximate treatment effects, e.g., the Gaussian process has shown potential as a surrogate for epidemic dynamics which could be used to estimate treatment effects \cite{ahmed2023inferring}.

\vspace{-3pt}
{\footnotesize
\section*{Data and Code Availability}
In this study, we are not able to share detailed data about the OUD model for contractual reasons. The repository link for the paper's code can be found at https://github.com/abdulrahmanfci/intervention-estimation. 

\section*{acknowledgment}
This research was funded by contract 75D30121C12574 from the Centers for Disease Control and Prevention. The findings and conclusions in this work are those of the authors and do not necessarily represent the official position of the Centers for Disease Control and Prevention.

This research was supported in part by the University of Pittsburgh Center for Research Computing, RRID:SCR\textunderscore022735, through the resources provided. Specifically, this work used the HTC and VIZ clusters, which are supported by NIH award number S10OD028483.
}

\bibliographystyle{IEEEtran}
\bibliography{bib.bib}

\end{document}